\documentclass[final,2p,times,twocolumn]{elsarticle}

\usepackage{amssymb}

\usepackage{graphicx}%
\usepackage{multirow}%
\usepackage{amsmath,amssymb,amsfonts}%
\usepackage{amsthm}%
\usepackage{mathrsfs}%
\usepackage[title]{appendix}%
\usepackage{xcolor}%
\usepackage{textcomp}%
\usepackage{manyfoot}%
\usepackage{array}
\usepackage{booktabs}%
\usepackage{algorithm}%
\usepackage{algorithmicx}%
\usepackage{algpseudocode}%
\usepackage{listings}%

\journal{ }

\begin{document}

\begin{frontmatter}



\title{Towards participatory multi-modeling for policy support across domains and scales: a systematic procedure for integral multi-model design.}


\author[inst1,inst2,inst3]{Vittorio Nespeca}
\author[inst1,inst2]{Rick Quax}
\author[inst4]{Marcel G. M. Olde Rikkert}
\author[inst5]{Hubert P. L. M. Korzilius}
\author[inst5]{Vincent A. W. J. Marchau}
\author[inst4]{Sophie Hadijsotiriou}
\author[inst4]{Tom Oreel}
\author[inst5]{Jannie Coenen}
\author[inst6]{Heiman Wertheim}
\author[inst7]{Alexey Voinov}
\author[inst5]{Etiënne A.J.A. Rouwette}
\author[inst1,inst2,inst8]{V\'itor V. Vasconcelos}

\affiliation[inst1]{organization={Computational Science Lab, University of Amsterdam},
            addressline={Science Park 900}, 
            city={Amsterdam},
            postcode={1098 XH}, 
            country={The Netherlands}}

\affiliation[inst2]{organization={POLDER, Institute for Advanced Study, University of Amsterdam},
            addressline={Oude Turfmarkt 147}, 
            city={Amsterdam},
            postcode={1012 GC}, 
            country={The Netherlands}}

\affiliation[inst3]{organization={Faculty of Technology Policy and Management, Delft University of Technology},
            addressline={Oude Turfmarkt 147}, 
            city={Delft},
            postcode={2628 BX}, 
            country={The Netherlands}}

\affiliation[inst4]{organization={Department Geriatrics, Radboud University Medical Center},
            addressline={Geert Grooteplein Zuid 10}, 
            city={Nijmegen},
            postcode={6525 GA}, 
            country={The Netherlands}}

\affiliation[inst5]{organization={Institute for Management Research, Radboud University},
            addressline={Heyendaalseweg 141}, 
            city={Nijmegen},
            postcode={6525 AJ}, 
            country={The Netherlands}}

\affiliation[inst6]{organization={Department Medical Microbiology, Radboud University Medical Center},
            addressline={Geert Grooteplein Zuid 10}, 
            city={Nijmegen},
            postcode={6525 GA}, 
            country={The Netherlands}}

\affiliation[inst7]{organization={Faculty of Engineering Technology, Twente University},
            city={Enschede},
            postcode={7522 NB}, 
            country={The Netherlands}}

\affiliation[inst8]{organization={Centre for Urban Mental Health, University of Amsterdam},
            city={Amsterdam},
            country={The Netherlands}}

\begin{abstract}
Policymaking for complex challenges such as pandemics necessitates the consideration of intricate implications across multiple domains and scales. Computational models can support policymaking, but a single model is often insufficient for such multidomain and scale challenges. Multi-models comprising several interacting computational models at different scales or relying on different modeling paradigms offer a potential solution. Such multi-models can be assembled from existing computational models (i.e., integrated modeling) or be designed conceptually as a whole before their computational implementation (i.e., integral modeling). Integral modeling is particularly valuable for novel policy problems, such as those faced in the early stages of a pandemic, where relevant models may be unavailable or lack standard documentation. Designing such multi-models through an integral approach is, however, a complex task requiring the collaboration of modelers and experts from various domains. In this collaborative effort, modelers must precisely define the domain knowledge needed from experts and establish a systematic procedure for translating such knowledge into a multi-model. Yet, these requirements and systematic procedures are currently lacking for multi-models that are both multiscale and multi-paradigm. We address this challenge by introducing a procedure for developing multi-models with an integral approach based on clearly defined domain knowledge requirements derived from literature. We illustrate this procedure using the case of school closure policies in the Netherlands during the COVID-19 pandemic, revealing their potential implications in the short and long term and across the healthcare and educational domains. The requirements and procedure provided in this article advance the application of integral multi-modeling for policy support in multiscale and multidomain contexts.
\end{abstract}




\end{frontmatter}

\section{Introduction}
\label{sec:introduction}
Designing policies in complex systems requires consideration that a given policy, designed for a single domain, may also have unintended effects in other domains \citep{klement_systems_2020, goenka2022smart}. For example, there is strong evidence that the school closure policies introduced to curb the spread of COVID-19 (healthcare domain) had prolonged adverse effects on the education of Dutch children (educational domain), especially for those belonging to the most economically disadvantaged families (social inequality domain) \citep{Haelermans_2022, engzell2021learning}. In addition to multidomain effects, policy implications can also span multiple time (e.g., in the short vs long term) and governance scales (e.g., central vs de-central \citep{janssen2020agile}). These are referred to as multiscale. For example, school closure policies (short term) led to prolonged learning loss for multiple years (medium term), and they may produce a negative effect on economic opportunities today’s children will have in the coming decades (long term) \citep{betthauser2023systematic, agostinelli2022great}.

Computational models can effectively map complex multidomain and multiscale implications and provide policy support \citep{gilbert2018computational,molina2023harnessing,smitsusing,araz2013simulation, barracosa2023decision,ertem2022decision}. For example, computational models have the potential to aid in the formulation of policies by exploring their implications across multiple domains and scales before their implementation \citep{gilbert2018computational}. However, as exemplified by the case of the COVID-19 pandemic, the models developed for policy support typically focused on one domain: the epidemiology of SARS-CoV-2 \citep{badham2021justified,ertem2022decision, ferguson2020report}. Moreover, these models often investigated the short-term implications of policy interventions without considering their potential long-term impacts \citep{badham2021justified, ferguson2020report, bicher2022supporting} or overlooked central/de-central dynamics even when these were relevant  \citep{janssen2020agile, dekker2023reducing}. One possible reason for such design choices may be that capturing multiscale and multidomain dynamics with one model can be difficult, if not impossible. 
For example, it would be computationally unfeasible to run simulations with one model capturing both the impacts of school closures on the epidemiology of SARS-CoV-2 (with a time step of days and run for a few years of a pandemic) and the long-term implications that such closures could have on the economic opportunities of children in the future (e.g., with a time step of years over a period of decades). Using a combination of models (or multi-models) instead of one model shows promise in addressing this challenge \citep{zeigler1986multifaceted, vangheluwe2002introduction, borgdorff2014performance, van_delden_integration_2007, fulton2015multi, bollinger2018multi, iwanaga2021socio}. 


Multimodels are models composed of two or more interacting sub-models capturing a phenomenon of interest from different perspectives including different domains (e.g., ecology and economics), scales, and modeling paradigms. Multimodels have been developed in several fields of research, including fluids and material science \citep{weinan_heterogeneous_2007, weinan2011principles}, biomedicine \citep{bellomo2020multiscale, sloot2010multi, wang2022multiscale}, and environmental science and engineering \citep{laniak2013integrated}. In addition to multimodels, these models are often termed multi-simulation, multi-resolution \citep{yilmaz2007requirements}, multi-scale \citep{weinan_heterogeneous_2007}, multi-paradigm, hybrid \citep{zeigler1986multifaceted, vangheluwe2002introduction}, or integrated modeling \citep{laniak2013integrated}. In multiscale (or multiresolution) modeling, the sub-models capture smaller portions of the original scales at which the dynamics of interest occur \citep{chopard2014framework}. For example, different models could be developed for the short- and long-term implications of particular policies (i.e., the implications at different time scales) or at the national and regional levels (i.e., implications at different governance levels). Such an approach inevitably entails approximations (e.g., a coarser level of detail captured only at the national level), but it can also lead to increased computational feasibility \citep{borgdorff2014performance}. Multi-paradigm (or hybrid) modeling focuses on capturing the subcomponents of a system through adequate modeling techniques or paradigms. Examples of such paradigms include Agent-Based Modeling (ABM) and System Dynamics Modelling (SDM). The choice of an adequate modeling paradigm can be informed by several criteria, including the purpose of the modeling endeavor, the scale at which key dynamics take place (e.g., at the aggregate level, or resulting from micro-level interactions), and the level of detail at which data and knowledge are available \citep{zeigler1986multifaceted,kelly2013selecting,scholl2001agent,Bonabeau2002,borshchev2004system}. When combined, multiscale and multi-paradigm modeling provide greater flexibility than individual models in allocating model complexity where required and feasible, enabling researchers to seek a balance among model complexity, the availability of knowledge and data, and computational costs \citep{kelly2013selecting, chopard2014framework}.

The design of such multi-models can rely on two strategies: integral and integrated modeling \citep{voinov2013integronsters}. Integral modeling entails designing the whole structure of a multi-model starting at the conceptual level \citep{voinov2020integrated}, followed by sub-model development. This approach ensures the ontological compatibility of the sub-models within the overall multi-model
(e.g., the World 3 model \citep{meadows1975limits}). Conversely, integrated modeling focuses on enabling the reuse of models developed by external parties. Integration occurs at the computational level, combining legacy models developed and implemented as software for other purposes \citep{laniak2013integrated, voinov2020integrated, tucker2022csdms}. While enabling model reuse, integrated modeling entails the risk of combining ontologically incompatible models (e.g., when they consider different categories of meaning or scales for the same variable) \citep{voinov2013integronsters}. All models are built for particular purposes; ensuring that reused models can effectively match the purposes driving the integrated model building is challenging. These risks are elevated in poorly documented models that do not follow joint standards \citep{squazzoni2020computational}.
Adopting an integral modeling approach is better suited than integrated modeling when decision-makers are presented with novel/unprecedented policy problems (e.g.,  the COVID-19 pandemic, the Syrian conflict and refugee crisis, or Brexit). In such cases, integrated modeling may limit the exploration of policy implications to the factors and dynamics captured in existing models, which, given the novelty of the policy problem considered, may still be lacking relevant domains and scales. Although models representing relevant domains and scales might exist, acknowledging their presence and establishing meaningful relationships among them demands a transdisciplinary perspective that may not be available to modelers (especially for novel policy problems). Conversely, an integral modeling approach enables one to explore the potential implications of policy across a broader range of domains and scales before implementing a multi-model. This is achieved first by constructing a conceptual and transdisciplinary understanding of the policy issue as a whole (possibly in collaboration with experts from multiple domains). Only then is the resulting conceptual model divided into interacting sub-models. This integral modeling design process results in a Multi-Modelling Structure (MMS) prescribing the sub-models in which the conceptual model is divided and how such sub-models interact. An MMS represents the ‘skeleton’ of a multi-model in that it defines the sub-models as ‘black boxes’ (i.e., without specifying their internal structure and software implementation) and represents the sub-models' scale, paradigms, and interactions (i.e., when and how the sub-models exchange information). The concept of an MMS is analogous to that of ‘architecture’ introduced by \citep{chopard2014framework} for multiscale modeling. One key difference is that an MMS specifies not only the scales associated with the sub-models but also their modeling paradigms. To complete the design of a multimodel given an MMS, modelers need to develop each of the sub-models, for example, through existing methodologies \citep{voinov2018tools, van_burggen_2019modeling} or employ (and adapt) existing models where these are available. Given that the sub-models’ interactions are already pre-defined in the MMS, the sub-models can be developed rather independently (e.g., by different teams).

Developing such an MMS representing multidomain and multiscale policy problems through an integral approach requires domain knowledge that spans several fields of expertise. As such, this knowledge may not be readily available to one modeler or even a team of modelers. Therefore, it seems paramount to co-create MMS with experts (e.g., decision-makers, their advisors, and stakeholders \citep{cuppen2021participatory}) from different domains of interest \citep{voinov2013integronsters,rouwette2002group,voinov2010modelling, buffa1996macrame}. In such a participatory process, a systematic and transparent approach is key at every step so that experts can understand the model, contribute to its development, trust the results to support policymaking, reach a consensus with respect to the policy interventions, understand and accept the uncertainties associated with the model, and accept and commit to implementing the decisions made through using the model \citep{vennix_group_1996, rouwette2002group, rouwette2011modeling, voinov2010modelling,voinov2016modelling,freebairn2017knowledge}. Ensuring such transparency requires clarity for modelers concerning (a) the domain knowledge needed from experts and (b) model development procedures that provide the means to translate this knowledge into a computational model. Participatory modeling techniques such as Group Model Building (GMB) clarify the co-development of single models (SDMs in the case of GMB) \citep{ford1998expert,mooy2001quantification,crielaard2022refining}. For example, Crielaard et al. \citep{crielaard2022refining} provide a series of requirements (or ‘annotations’) that capture the domain knowledge required from experts and a systematic and transparent procedure to translate this knowledge into an SDM. 

In the case of the design of MMSs with an integral approach, several requirements and procedures have been proposed for multiscale \citep{chopard2014framework} or multi-paradigm modeling \citep{chahal2013conceptual, mykoniatis2020modeling, deLara2004meta, muller2004mimosa, muller2010framework} without however combining these two perspectives. Kelly et al. \citep{kelly2013selecting} guide the choice among different modeling paradigms, including considerations about the required level of spatial detail (or scale). Such guidelines provide the domain knowledge required to choose among different modeling paradigms, however, they do not capture the additional domain knowledge required to functionally decompose the system of interest in subcomponents at different scales, which is essential for developing an MMS. Clearly defined domain knowledge requirements and a procedure to systematically and transparently develop MMSs with experts in multiscale and multidomain contexts are missing. 

This article proposes a systematic procedure to develop MMSs with an integral approach based on clearly defined domain knowledge requirements derived from multi-paradigm and multiscale modeling literature. We illustrate this procedure using the case of school closures in the Netherlands by looking at their multiscale and multidomain implications, which cannot be understood when considering a single domain and scale in isolation. 

\section*{Design of the Procedure}\label{results}
In this section, the procedure for developing MMSs designed based on requirements derived from literature as shown in the following sections. 

\subsection{Requirements for designing multi-modeling architectures}
We identify domain knowledge requirements for MMSs by analyzing literature from multiscale modeling and multi-paradigm modeling. That is, for each of these two fields, we derive a list of key design choices to be made by modelers when developing multimodels. Then, we infer the domain knowledge required from experts to make these modeling design choices informedly, which we refer to as the domain knowledge requirements. The details of this process are in the Appendix, while the results are shown in Table \ref{tab:req-multi-modeling}. Precise definitions for the design choices and requirements shown in Table \ref{tab:req-multi-modeling} are provided in the following sections. 

\begin{table*}[h!]
\caption{Key choices in the design of Multi-Model Structures (MMSs), and the domain knowledge requirements for each of the choices with some clarifying examples. Some requirements, such as ‘Policy problem,’ are repeated in the right column as they can be required for more than one of the design choices. Repeated requirements do not include clarifying examples.}\label{tab:req-multi-modeling}
\begin{tabular*}{\textwidth}{m{0.3\textwidth} m{0.64\textwidth}}
\toprule%
\textbf{Key design choices} & \textbf{Domain knowledge requirements} \\ \midrule
Identification of policy problem and modeling objective & 
 Policy problem (e.g., the formulation of school closure policies to tackle a pandemic), Constraints (e.g., budget or time)\\ \midrule
Functional decomposition in sub-models 1: Scale separation (or decomposition) & 
Scale dimensions (e.g., time, space, or governance level), Scale intervals within the scale dimensions (e.g., a scale interval of years to decades in the time scale dimension)
\\ \midrule
Functional decomposition in sub-models 2: Modeling paradigm selection &  Policy problem, Constraints, Scale intervals, Heterogeneity (e.g., diversity in immune response in the considered population), Mixing (e.g., patterns of interactions among individuals and their environment), Complexity of individual behavior (e.g., learning or adaptation), Discreteness of system’s behavior (e.g., a disease is eradicated when there are less than one person infected or susceptible), Uncertainty in and availability of data and knowledge (e.g., the level of detail and precision at which data and knowledge are available in the considered system).
\\ \midrule
Definition of information exchange interactions among sub-models & Functional decomposition in sub-models 1 \& 2, Time scale intervals: informs coupling temples (when models exchange information, e.g., during or before/after the simulation, how often, what is the time-stepping), Scale intervals in dimensions other than time (e.g., structural, say, governance or spatial): informs the necessity for scale bridging operations such as aggregation or disaggregation, Topology of interactions (which sub-model provides info. to which sub-model),  Operations (e.g., equations, if-then-else statements, aggregation/disaggregation techniques to process information exchanged across interacting models).\\
\bottomrule
\end{tabular*}
\end{table*}

\subsection*{Procedure for developing multimodeling structures based on the requirements} \label{results_theoretical}

We draw the procedure for developing MMSs from the previous section's design choices and domain knowledge requirements (cf. Table \ref{tab:req-multi-modeling}). Specifically, we organize the steps of the procedure in terms of the dependencies among requirements and design choices, from the most fundamental — i.e., the requirements and design choices that directly or indirectly affect most of the following design choices — to those that are least fundamental and, as such, depend on other design choices and requirements. Additionally, the model development process is centered on and begins with the design of a conceptual model according to the integral modeling approach selected in this article for novel policy problems. 

First, the most fundamental requirements are the policy problem and project constraints. These enable modelers to define the modeling objective (crucial throughout the multi-model development process). Further, the policy problem and project constraints also inform the choice of modeling paradigms. As such, these two requirements must first be elicited to inform the choice of a modeling objective. Given the modeling objective, the development of a conceptual model begins. Casual mapping techniques, such as Cognitive Maps and Causal Loop Diagrams (CLDs), are effective for conceptual modeling \citep{barbrook2022systems}. Here, we focus on the CLD as the conceptual modeling framework. CLDs are a graphical tool to develop a qualitative understanding of the ‘structure’ of a system in terms of its relevant factors (concepts) and the causal relationships among them. This structure can provide qualitative insights regarding the functioning of the system, e.g., by providing a preliminary understanding of mechanisms (e.g., feedback loops, reinforcing and balancing) that drive and explain system behavior \citep{sterman2000business}. The factors and relationships in a CLD provide a conceptual framework that facilitates the elicitation and graphical representation of the additional expert knowledge required to translate CLDs into computational models \citep{ford1998expert,crielaard2022refining} (in this case those shown in Table \ref{tab:req-multi-modeling}), making it a natural starting point for this procedure. This first step, defining the policy problem, constraints, modeling objective, and causal model, is termed \textit{Conceptual Modeling}.

The second most fundamental requirements are the scale dimensions and intervals. This is the case as such requirements are necessary for all of the remaining design choices namely, the scale decomposition (or separation), selection of the modeling paradigms, and design of the interactions among the sub-models. These three design choices also present interdependencies as the scale decomposition is required to inform the paradigm selection, and both the scale and paradigm decomposition are required to define the interactions. Then, scale separation constitutes the most fundamental among the three design choices, given it needs to be carried out before the other two are executed. As such, the second step of the procedure focuses on (a) defining the scale dimensions and intervals, and (b) carrying out the scale separation based on such intervals and dimensions. This step is called \textit{scale decomposition}, and its results are a preliminary division in sub-models at different scales and the identification of (causal) relationships among factors belonging to different sub-models. 

Next, the next most fundamental requirements are those necessary to inform the selection of the modeling paradigms (e.g., heterogeneity, discreteness in system behavior, and availability of data and knowledge cf. Table \ref{tab:req-multi-modeling}). Here, not only are modeling paradigms assigned to the sub-models obtained through the scale decomposition but also such decomposition can be revisited, e.g., by further disaggregating the sub-models or by aggregating them depending on the modeling paradigm adopted. This step, termed paradigm decomposition, results in the final sub-division in sub-models with the associated scale intervals and modeling paradigms. 

\begin{figure} [htp]
    \centering
    \includegraphics[width=0.41\textwidth]{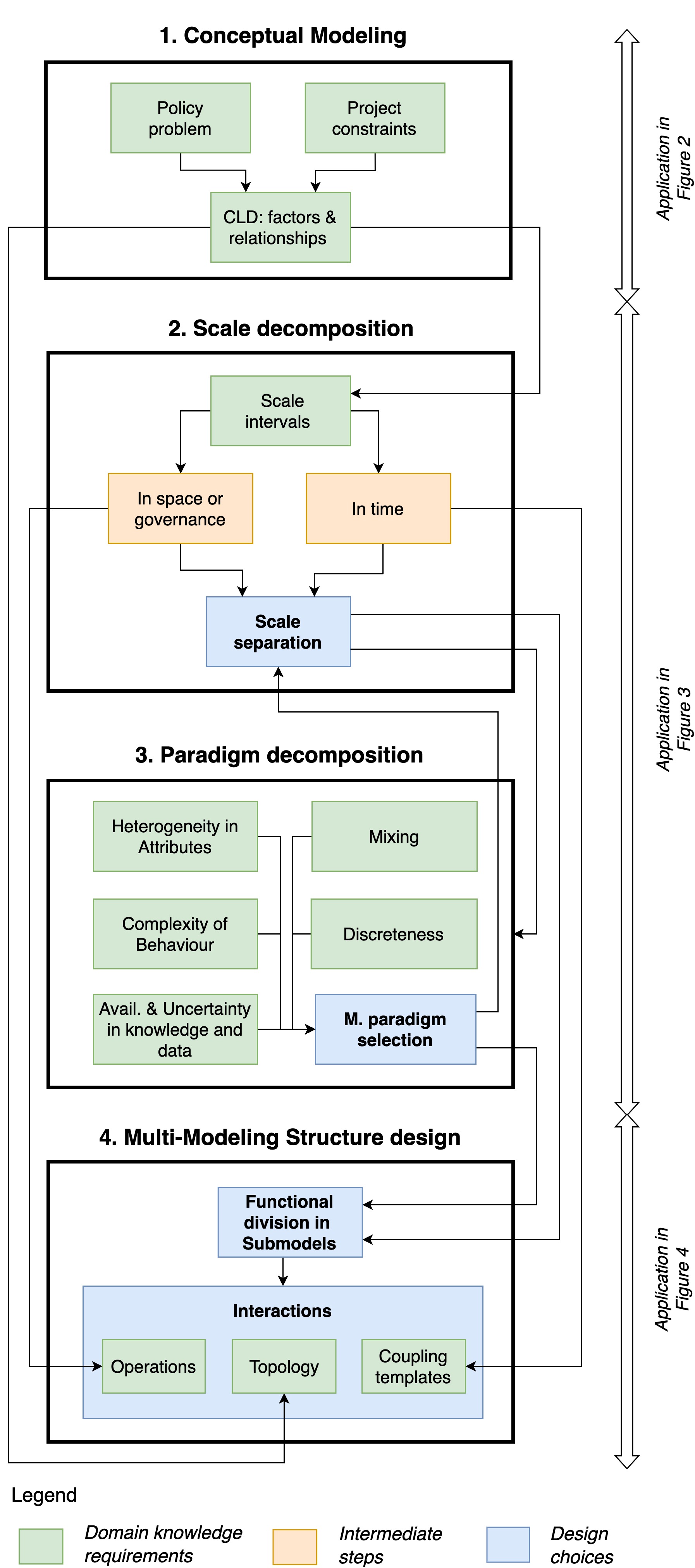} 
    \caption{Procedure for developing Multi-Model Structures (MMSs). First, develop a Causal Loop Diagram (CLD) that captures the relevant factors and relationships for the considered system and policy problem. Then, given the scale intervals associated with the factors, functionally decompose the CLD in sub-models at different scales (e.g., in time and governance level). Next, functionally decompose the sub-models by assigning adequate modeling paradigms informed by additional domain knowledge on, e.g., heterogeneity and discreteness in the system’s behavior. Finally, given the CLD and the functional decomposition in scale and paradigms, design an MMS consisting of the submodels and their interactions (topology of interactions, operations, and coupling templates).}
    \label{fig:procedure}
\end{figure}

Finally, the last requirements to be elicited are those necessary to inform the definition of the interactions among the sub-models. Most of these requirements are already available from the previous steps (e.g., scale intervals). However, the operations carried out on information when sub-models exchange information are partly missing. This is because, while aggregation/disaggregation operations can be inferred from the scale intervals (e.g., in space or governance level), other operations, such as mathematical expressions representing causal relationships or logical statements, are yet to be defined. As such, the last step of the procedure focuses on defining the operations carried out on information when the sub-models exchange information. Once these requirements are fulfilled, the modeler can make the last design choice, which is to fully define the interactions among the sub-models (composed of the topology of interactions, operations, and coupling templates). Given the sub-models (including their scales and paradigms) and their interactions, the modeler can assemble a Multi-Modeling structure (MMS). As such, the last step of the procedure is termed \textit{Multi-Modeling Structure design}. The procedure with all of its steps is shown in Figure \ref{fig:procedure}.

Once the MMS is developed, the modelers can rely on other existing methodologies to develop each sub-model (or re-use existing models).  This last step is not included in the procedure as previously existing methodologies can be applied to develop the sub-models \citep{voinov2018tools, van_burggen_2019modeling}. In the following sections, we describe each of the steps of the procedure in theory and illustrate their application in practice through the case of School Closures in the Netherlands (introduced hereinafter).

\subsection{Case Study: School Closures in the Netherlands} \label{results_application}

The closure of schools in the Netherlands is an opportune case for illustrating the procedure, given that, as discussed in the introduction, it presents policy implications and dynamics that are both multidomain and multiscale. First, multiple domains are affected by school closures. For illustration, we focus on healthcare and education \citep{araz2013simulation}. In the healthcare domain, school closures were introduced to curb the spread of COVID-19, producing results that are still debated \citep{el_jaouhari_impact_2021, walsh_school_2021,krishnaratne_measures_2022}. In the education domain, school closures caused learning loss for children, especially for those from the most economically disadvantaged families \citep{Haelermans_2022,engzell2021learning}. Second, the implications of school closures span multiple time scales as the learning loss accumulated by children in the short term is expected to have long-term effects on their economic opportunities in the future \citep{agostinelli2022great}. We illustrate the steps of the procedure to develop an MMS (cf. Figure \ref{fig:procedure}) based on this case study as follows. We consider 1000 individuals (agents) to represent the Netherlands and, as such, consider average characteristics (e.g., percentage of highly skilled pupils) equal to those at the national level. The goal of this application is not to develop an MMS capturing the implications of school closures accurately. Rather, the aim is to show that, given the required domain knowledge (Cf. Table 1), the proposed procedure enables modelers to develop an MMS that captures policy implications spanning multiple scales and domains.

\subsubsection{Step 1: Conceptual Modeling}
\textbf{\textit{Theory}}: Much research has focused on developing CLDs based on literature, participatory processes, quantitative data, and combinations of these approaches. As such, we refer the reader to previous work in CLD development such as \citep{vennix_group_1996, yearworth2013uses, wittenborn2016depression,uleman2021mapping} to learn more about how to carry out this step.

\begin{figure*} [h!]
    \centering
    \includegraphics[width=\textwidth]{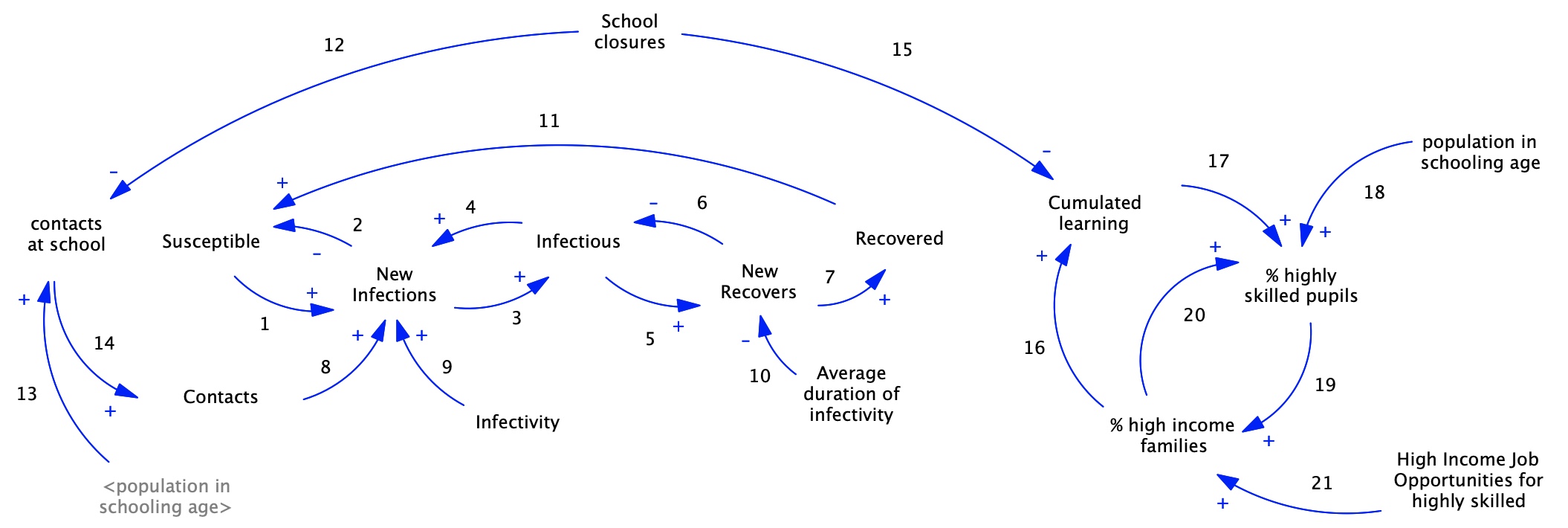} 
    \caption{Conceptual Modeling: the Causal Loop Diagram (CLD) captures the implications of school closures for healthcare and education through a series of factors and relationships among them. Each relationship is assigned a number from 1 to 21 to facilitate their discussion. For healthcare, the core of this CLD is a Susceptible Infected and Recovered (SIR) model (1 - 10) \citep{kermack1927contribution} with the addition of immunity loss (11) \citep{andeweg2022protection}. School closures reduce contacts at school (12) among the school-age population (13), which contributes to the general number of contacts (14). For education, the CLD includes the negative effect of school closures on the cumulated learning of the school-age population (15) \citep{engzell2021learning}. Such an effect is reduced by the percentage of the children’s families with a high income (16) \citep{engzell2021learning}. Further, higher cumulated learning increases the percentage of highly skilled pupils (17) who have higher chances of obtaining high-income jobs (19 \& 21), resulting in a higher percentage of high-income families \citep{oecd_education_2022}. The percentage of high-income families improves the education opportunities of children from such families, thus increasing the percentage of highly skilled pupils (20).}
    \label{fig:cld}
\end{figure*}

\textit{\textbf{Application:}} In this case, the development of a CLD capturing the implications of school closures for healthcare and education was based on literature, as shown in Figure \ref{fig:cld}.

\subsubsection{Step 2: Scale Decomposition}
\textbf{\textit{Theory:}} At this stage, the CLD from the previous step is divided into sub-models at smaller scale intervals for each of the considered dimensions (e.g., time, space, and governance level). Given that the purpose here is to subdivide the considered system into sub-systems to be described by different models, this CLD is intended as a ‘bird’s eye’ view of the factors and relationships relevant to the considered system and policy problem. The scale decomposition proceeds in two steps. Firstly, scale intervals are assigned to each factor in the CLD for each dimension considered. This step includes defining the relevant dimensions along which scale decomposition is carried out. Then, scale intervals are assigned to the factors along the considered dimensions.

A scale interval is composed of two values. Both values are not precise numbers but, rather, they represent orders of magnitude. The first value represents the granularity with which it is relevant to appreciate a given factor in a given dimension. The second value represents the total size that a given factor has for the considered dimension \citep{borgdorff2013foundations}. 

For example, in the case of \textit{time scale intervals}, the first value is the frequency with which it is relevant to assess changes in the considered factor (e.g., “days”); it corresponds to the model time step needed to capture the factor’s dynamics. The second value is the time horizon over which it is relevant to study changes in the factor (e.g., ”months”). This value corresponds to the necessary duration of the simulations (accumulation of time steps) to study the factor. As such, a time scale interval of ”(days, months)” means that the considered factor can be simulated with a time step of days over a period of months.
In the case of \textit{governance scale intervals}, the first value represents the governance level at which a factor is determined. The second value represents the governance level at which the factor is relevant for decision-making. As such, a governance scale interval of (“city”, “regional”) suggests that a particular factor is determined at the city level and has implications for decision-making at the regional level. Depending on the focus of the policy problem considered, there may be different levels of decision-making relevant to the considered system, e.g., at the municipality, regional, and national levels. As such, the same factor may have varying governance intervals depending on stakeholders' perspectives at different governance levels. When this occurs, the factor should be subdivided into different factors, each with a unique governance scale interval. This is especially relevant when the same factor has different implications for decision-making at different governance levels.

Secondly, once the scale intervals have been assigned, the factors are divided into groups with homogenous scale intervals in the considered dimensions. Each group constitutes a sub-model resulting from the scale decomposition. Further, the relationships that connect factors belonging to different groups represent the interactions among the sub-models. In the following, we discuss these two steps in detail through the case of school closures.

\textbf{\textit{Application:}} The dimensions considered relevant for the case of school closures are governance and time. As such, scale intervals were assigned to each factor for both the governance and time dimensions. With regards to the governance scale intervals for the model presented in Figure 2, most factors were found to have a governance scale interval of (“Local”, ”National”). For example, the factor ”new infections” is determined through contacts among susceptible and infected at the local level. As such, the first value corresponds to the local level. Regarding the second value, ”new infections” is observed here at the national scale, given that decision-making for this illustrative application occurs at the national level. Different decision-making levels may be relevant in the considered case, e.g., when modeling central / de-central dynamics between the regional and national governance levels \citep{janssen2020agile}. However, for simplicity, we only considered the national decision-making level. The “school closure” factor had a different scale interval as it is determined at the national level and has implications at the national level. As such, for this factor, the scale interval is defined as (“National”, ”National”). Other factors were assigned the same governance scale interval as they represent characteristics of the system at the National level. This was the case for “Population in schooling age” and “High income job opportunities.” Finally, some of the factors were unrelated to any governance level, so a governance scale interval was not assigned. This was the case for the epidemiological factors “infectivity” and “average duration of infectivity” which represent exogenous characteristics of the COVID-19 virus for which the governance dimension is irrelevant.

Next, the time scale intervals were assigned. In terms of the first value of the scale, several values were found for different factors, ranging from the daily to the yearly frequency. Regarding the second value, or time horizon over which it is relevant to study a given factor, the factors in the CLD ranged from years to decades. For example, the factor ‘cumulated learning’ was found to have a scale interval of (“days”, ”decades”). The first value was set to “days” as kids learn at school daily, so this is the frequency chosen to assess learning. Further, the learning process for pupils who are preparing for university or the job market takes several years, as such, the time horizon considered for this factor is that of years. “Infectivity” and “average duration of infectivity” do not change over time. As such, the scale interval for these factors is set to (“Constant”).

\begin{figure*} [htp]
    \centering
    \includegraphics[width=\textwidth]{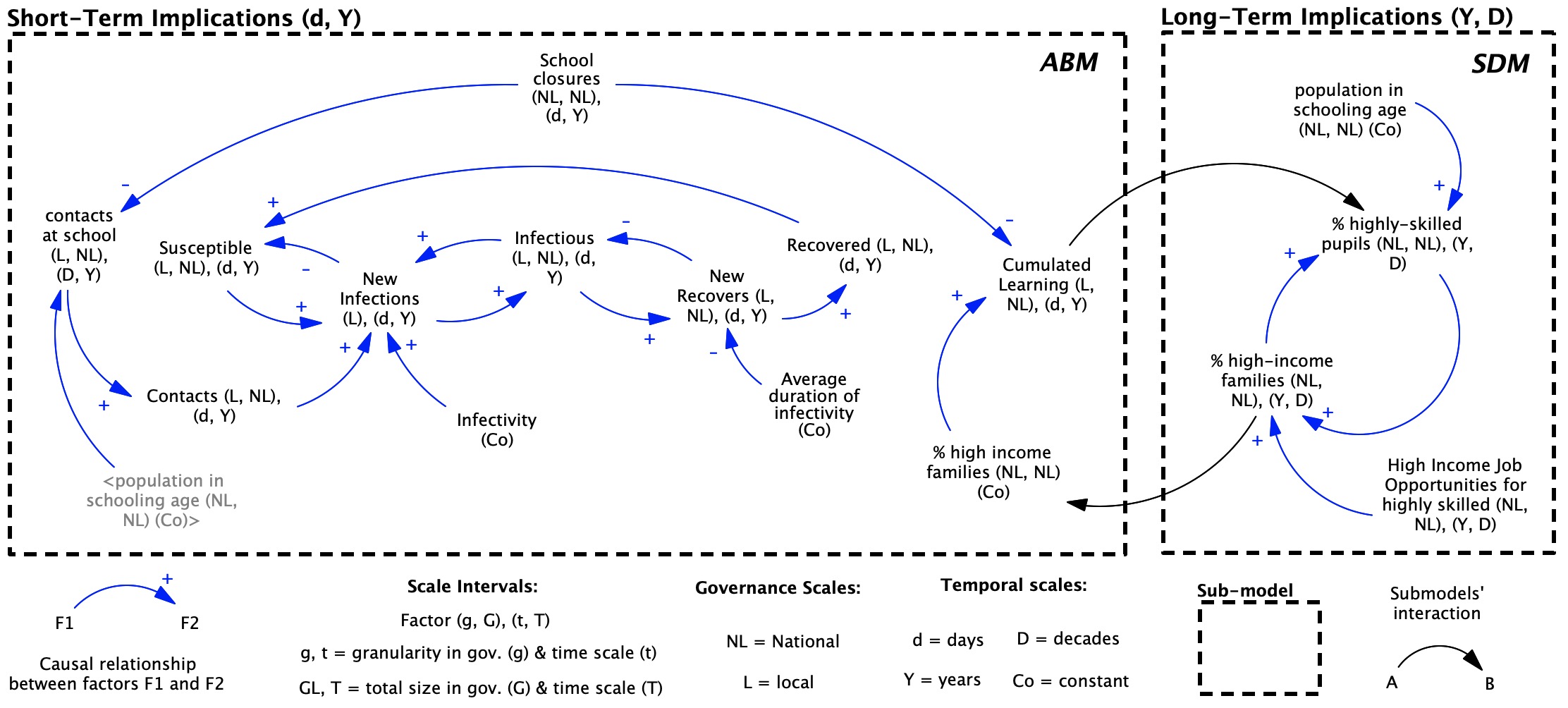} 
    \caption{Scale and Paradigm Decomposition: The Causal Loop Diagram (CLD) obtained through Conceptual Modeling (factors and causal relationships among them) is decomposed into sub-models at different temporal scales (short- and long-term) and relying on different modeling paradigms (AgentBased Modeling (ABM) and System Dynamics Modeling (SDM)).}
    \label{fig:scale-and-paradigm-separated-cld}
\end{figure*}

Then, the factors are divided into groups characterized by homogeneous intervals in time and governance scales. In terms of time or temporal scales, the factors are organized into two macro groups: the short-term, with scale intervals of days to years, and the long-term, with scale intervals from years to decades. Regarding the governance-scale intervals, no functional decomposition is required as all factors in the model present the same interval, ranging from local to national level. The time and governance scales assigned to the factors and their scale decomposition in submodels are shown in Figure \ref{fig:scale-and-paradigm-separated-cld}. Additionally, the scale decomposition enabled the identification of relationships among factors that belong to different sub-models, representing the topology of interactions among the sub-models. Specifically, the interactions found are two (cf. Figures \ref{fig:scale-and-paradigm-separated-cld}). First, the relationship between “Cumulated Learning” and “\% highly-skilled pupils” shows that the short-term model influences the long-term model through the effect that cumulated learning has on the fraction of the pupils (approaching university or the job market) that are highly skilled. Further, the relationship between the “\% high-income families” and “Cumulated Learning” shows how the long-term model affects the short-term model. At this stage, an additional factor was added to the CLD that replicates “\% high-income families” but with a constant value (Co) instead of a time scale interval ranging from years to decades (Years, Decades) (Cf. Figures \ref{fig:cld} and \ref{fig:scale-and-paradigm-separated-cld}). This additional value is added to illustrate a value transfer from the factor “\% high-income families (National, National) (Years, Decades)” which is dynamically changing every year in the long-term model, to the factors “\% high-income families (National, National)  (Constant)” which is set as a constant in the short term model based on the final result obtained by the long-term model.

\subsubsection{Step 3: Paradigm Decomposition}
\textbf{\textit{Theory:}} In this section, modeling paradigms are chosen for the sub-models identified during the scale decomposition. A substantial body of literature exists that focuses on the selection of a model paradigm aligned with specific criteria or domain knowledge, e.g., \citep{lorenz2006towards, chahal2013conceptual, mykoniatis2020modeling, kelly2013selecting}. It is not our intention in this article to provide a comprehensive list of domain knowledge requirements based on an exhaustive review of the existing literature. Instead, we aim to offer select illustrations of different requirements that can be used to choose a modeling paradigm for the sub-models obtained through scale separation. These requirements are the policy problem considered, project constraints (e.g., time and budget availability), characteristics of the considered system (e.g., heterogeneity and mixing), the availability of data and knowledge, and the uncertainties associated with such data and knowledge (Cf. Table \ref{tab:req-multi-modeling}). At this stage, the scale separation can also be revisited. For example, this can occur when an ABM is selected to capture two or more sub-models at different governance scale intervals in one single model. 

\textbf{Application:} In the following, the choice of a modeling paradigm is discussed for each sub-model obtained through scale decomposition, namely, the short and long-term implications of school closures.

In the case of the \textit{short-term implications} of school closures, the choice of a modeling paradigm was guided by the properties of the system associated with the contacts among susceptible and infected at school and elsewhere. With regard to contacts at school, and in the case of an airborne virus (especially if aerosols play a significant role), it can be assumed that every student has an equal chance of coming into contact with any other student in the same classroom. Such an opportunity for contact among all pupils corresponds to a perfect mixing hypothesis \citep{kermack1927contribution}. We extend this assumption to all schools in the considered system, thus assuming that all pupils attend the same classroom in the same school. It is imperative to acknowledge that this represents a notable simplification, as the actual system typically comprises multiple distinct classrooms and schools, thus demonstrating a lesser degree of perfect mixing. Consequently, such a system inherently presents a heightened potential for school-based infections compared to real-world scenarios. Then, any conclusions derived from this assumption are inclined to overstate the impact of school closures once contact among all pupils at school is stopped. Should school closures prove ineffective in such an idealized system, they are likely to be ineffective in a more realistic system, which accounts for less perfect mixing and, thus, fewer chances for infections among pupils at school.

In the case of contacts outside schools, a more accurate approximation entails that such contacts are not perfectly mixed and, rather, depend on networked interactions among individuals (e.g., based on their social networks) \citep{rahmandad2008heterogeneity}.

Further, the behavior of a system associated with eradicating an infectious disease is discrete. Continuous models, such as SDMs, do not capture eradication or extinction dynamics naturally, requiring arbitrary thresholds to stop the dynamics. For example, in the case of a closed system, since continuous models allow for less than one infected person (e.g., if the number of infected is 0.1), a disease can regrow as the number of susceptible individuals increases. A similar phenomenon occurs for the extinction of biological species \citep{wilson1998resolving}. Conversely, in the case of a discrete model of a closed system, when the number of infected (or susceptible) is less than one, the disease is eradicated. These systems’ characteristics indicated that an ABM was a reasonable choice given this modeling paradigm’s ability to capture non-perfect mixing and discreteness of the system behavior \citep{Bonabeau2002,dyke1998agent}. Further, the ABM modeling approach was known to the authors, and data could be retrieved for this example application from information gathered during the COVID-19 pandemic and other baseline sources.

With regards to the long-term implications of school closures, the choice of a modeling paradigm was mainly influenced by the unavailability of data on dynamics (e.g., ‘percentage of highly skilled pupils’ cf. Figure \ref{fig:scale-and-paradigm-separated-cld}) that did not occur yet and are, as such, highly uncertain. In this case, using an ABM would have required introducing additional details at the micro level for which data was unavailable, resulting in more uncertainty in the results \citep{rahmandad2008heterogeneity}. In this case, using an SDM focusing on the aggregate (macro) level was the preferred choice. Additionally, choosing an SDM model presents the advantage of reducing the computational costs associated with the long-term sub-model. As such, an SDM modeling paradigm was chosen for this sub-model. The resulting functional decomposition in scales and modeling paradigms is shown in Figure \ref{fig:scale-and-paradigm-separated-cld}.

\subsubsection{Step 4: Multi-modelling structure design}
\textbf{\textit{Theory:}} Designing an MMS entails defining the sub-models and their interactions. Such interactions represent information flows among models and comprise the topology of interactions, the operations carried out on the information exchanged among sub-models, and the coupling templates. First, the topology of interactions defines the structure and flow of information among submodels, and it is determined by the relationships that connect factors belonging to different sub-models obtained through the scale and paradigm decomposition steps.

Second, the operations to be carried out to process information when it is exchanged among different models are defined. The operations can include value transfer (without any additional operation), (dis)aggregation, mathematical operations, logical operations (if, then, else statements), or combinations of them.

Finally, the coupling templates establishing when the sub-models interact must be defined based on comparing the time scale intervals associated with the sub-models \citep{borgdorff2013foundations}. Scale intervals of different models can be separated, contiguous, or overlap. When the sub-model scale intervals overlap, the models run in parallel and exchange information during each simulation step. When the time scales of the sub-models do not overlap or are contiguous, then the models run in series \citep{chopard2014framework}. 

The topology of interactions, operations, coupling templates, and the submodels obtained through scale and paradigm separation constitute an MMS. Such an MMS can then be captured through the Multiscale Modelling Language (MML) proposed in \citep{falcone2010mml, borgdorff2013foundations, chopard2014framework}.

\textbf{\textit{Application:}} 
First, we identify the topology of the interactions. This topology was obtained in the scale decomposition (cf. ”Sub-models’ interactions” in Figure \ref{fig:scale-and-paradigm-separated-cld}). In our MMS, the short-term sub-model provides the long-term sub-model with ”cumulated learning”, affecting the ”\% highly skilled pupils”. Conversely, the long-term sub-model provides the short-term sub-model with the ”\% high-income families”.

Second, the operations to be carried out on the information exchanged among submodels were identified. This requires clarity on the interaction from the short-term submodel to the long-term sub-model (and back), requiring an operation as the value of ”cumulated learning” needs to be translated into the ”\% highly skilled pupils” (cf. Figure \ref{fig:scale-and-paradigm-separated-cld}). Particularly, the ”\% highly skilled pupils” is calculated by multiplying the percentage of highly skilled pupils before the pandemic by the “cumulated learning” during the pandemic, divided by the cumulated learning that pupils would have obtained in the absence of the pandemic (assuming no changes in the historical trends). In the case of the interaction from the long-term sub-model to the short-term sub-model, no additional operation is required as the value of ”\% high income families,” is directly transferred from the SDM to the ABM (cf. Figure \ref{fig:scale-and-paradigm-separated-cld}).

\begin{figure*} [h!]
    \centering
    \includegraphics[width=0.8\textwidth]{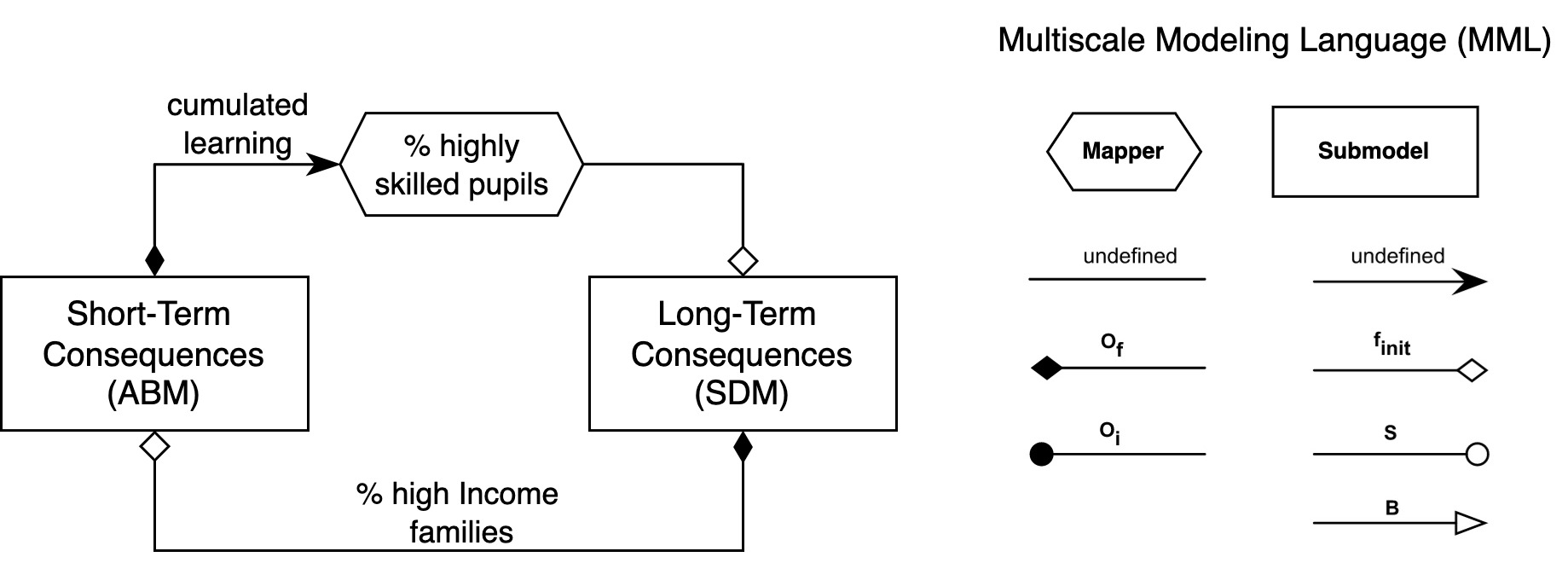} 
    \caption{
    Multi-Modelling Structure (MMS): the MMS (left) captured through the Multiscale Modeling Language or MML (right). The MMS shows the way the identified short- and long-term sub-models (cf. Figure \ref{fig:scale-and-paradigm-separated-cld}) exchange information and how such information is processed through operations (marked by mappers). For the MML, we refer the reader to Figure \ref{fig:mml} in the Appendix.
    }
    \label{fig:application-mml}
\end{figure*}

Finally, we established the coupling templates. The short-term sub-model has a time scale interval of days to years, whereas the long-term sub-model presents a time scale interval of years to decades. Given that the time scale intervals of the two submodels are contiguous, these models run in series. 
Specifically, the ABM is run with a time step of one day for two years (e.g., corresponding to the duration of the pandemic). Once completed, the resulting cumulated learning can be processed to obtain the initial value for the percentage of highly skilled pupils in the SDM model. Next, the SDM model is run for several decades. The simulation can be stopped here. Or, if another hypothetical pandemic is considered as a future scenario, the final value of ”\% high-income families” from the SDM can be passed onto the ABM to set the initial (and constant) value for ”\% high-income families.” The topology of interactions, operations, and coupling of the models are illustrated through the MML, resulting in the MMS shown in Figure \ref{fig:application-mml}. 

It must be stressed that another hypothetical pandemic is not considered as a future scenario for the sake of this illustrative study. However, it would be possible to account for such a scenario. Further, the Multiscale Modeling and Simulation Framework (MMSF) introduced by \citep{chopard2014framework} and presented in the Appendix facilitates the coupling of models running both in series and/or in parallel.

\subsubsection{Use of the Multi-Modeling Structure for Policy Formulation} 

\begin{figure*} [htp]
    \centering
    \includegraphics[width=\textwidth]{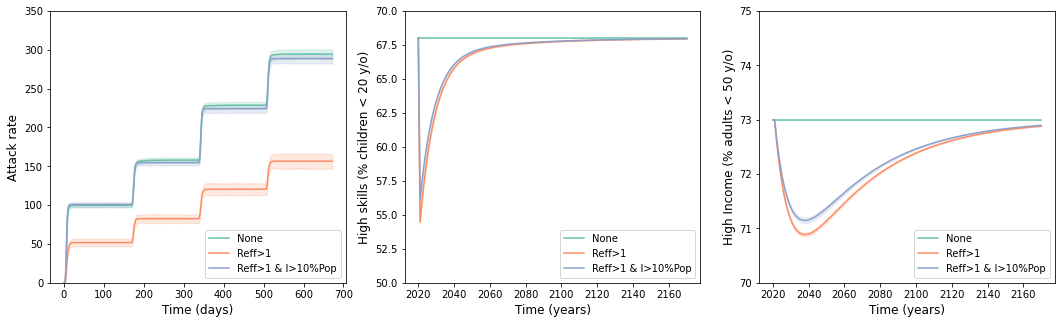} 
    \caption{Short- (left) and Long-term (middle and right) implications of school closure policies in the Netherlands. Three policies are considered. ”None” means that the school remains open at all times. ”Reff\textgreater1” means that schools are closed when the effective reproduction number is above one (assuming perfect information and no delays). ”Reff\textgreater1 \& I\textgreater10\%” is similar to the above, with the additional condition that at least 10 percent of the population becomes infected before closing schools. In the short term, policy effectiveness is measured in terms of the attack rate, expressed as the average number of times a person was infected during the pandemic. In the long term, policy implications are indicated by the “\% of highly skilled pupils” (individuals younger than 20 years old that have a higher chance of finding highly paid jobs) and by the “\% High Income” (percentage of the population older than 20 and younger than 50 that have a high income). For this illustrative example, it is assumed that no other exogenous factors (e.g., population or availability of highly skilled jobs) change over time throughout the simulation.}
    \label{fig:long-and-short-term-results}
\end{figure*}

Here, a multi-model was developed based on the MMS (cf. Figure \ref{fig:application-mml}) to simulate the implications of school closures across healthcare and education in the short- and long-term. This process entailed the development of the two sub-models (ABM and SDM) based on the identified factors and relationships included in the sub-models (cf. Figure \ref{fig:scale-and-paradigm-separated-cld}). Then, the sub-models and their interactions were implemented as software. Finally, simulation experiments were run to explore policy implications. The experimental results are shown in Figure \ref{fig:long-and-short-term-results}.

The results in Figure \ref{fig:long-and-short-term-results} (left) show that school closure policies effectively reduce the attack rate (i.e.,  the number of infections that a person was on average subject to throughout the pandemic) in the short term only in the unlikely case where the closure takes place immediately after Reff increases above 1 (policy ”Reff\(>\)1”). Even waiting for 10 percent of the population to be infected before closing schools makes the closures practically ineffective (policy ”Reff\(>\)1 \& I\(>\)10\%”). The inefficacy of school closures is particularly significant given that the perfect mixing assumption in the model overstates the opportunity for contacts at school (cf. Section ”Step 3: Paradigm Decomposition”). Assuming calibration happens at the contact level, if school closures are ineffective in such a system, they are likely to be even less effective in a system that, more realistically, presents fewer school contact opportunities.

The long-term negative effects on the skill level of pupils (cf. Figure \ref{fig:long-and-short-term-results} middle) and their opportunities for high income (cf. Figure 5 right) are consistently non-negligible for all school closure policies (”Reff\(>\)1” and ”Reff\(>\)1 \& I\(>\)10\%”) compared to the case in which no policy is introduced (“None”). A comparison of the inefficacy of school closures in reducing the attack rate in the short term with their long-term implications on acquired skills and incomes suggests that closure policies are not a very effective solution to implement \textit{in the context of this illustrative example}. As such, other possible policies could be considered as an alternative to and or in combination with school closures.

\section*{Discussion \& Conclusions}\label{methods}
We proposed a procedure for developing Multi-Model Structures (MMSs) for policy support in multiscale and domain contexts. This procedure was designed based on clearly defined expert knowledge requirements derived from the literature in multiscale and multi-paradigm modeling. We adopt an integral modeling approach, where the design of an MMS starts at the conceptual level 
\citep{voinov2013integronsters}. We illustrate this procedure through the case of COVID-19 school closures in the Netherlands to show how the procedure enables the design of MMSs systematically and transparently, clearly indicating what domain knowledge is required at each step and how the results of the previous steps inform the following steps of the procedure.

Implementing the designed MMS enabled exploring the implications of school closure policies across multiple time scales (the short and long term) and domains (healthcare and education). This result suggests that the proposed procedure can result in MMSs that support policy formulation in multiscale and multidomain contexts. 

This procedure has significant implications for developing transparent multi-models supporting policy-making across various scales and domains. It elucidates the domain knowledge essential for designing Multi-Model Structures (MMSs), establishing a foundational framework for creating participatory methods or 'scripts.' These scripts (cf. \citep{hovmand2012group})  facilitate the collaborative elicitation of domain-specific knowledge requirements with experts. This approach fosters the systematic and transparent co-creation of MMSs, enhancing policy formulation across diverse domains and scales.

The outlined procedure clarifies the initial step in multi-model development, pinpointing the requisite sub-models and their interplays to form an MMS. The subsequent phase involves the actual construction of these sub-models, where existing methodologies, such as those proposed by Ford and Sterman \citep{ford1998expert}, Mooy et al. \citep{mooy2001quantification}, and Crielaard et al. \citep{crielaard2022refining} can be effectively integrated. Thus, this process not only aids in crafting multi-models but also leverages established methodologies designed for individual models.

Furthermore, this procedure yields Casual Loop Diagrams (CLDs) within each sub-model as a beneficial by-product. These CLDs, encompassing various factors and relationships, can guide the development of the sub-models, aligning with the methodology proposed by \citep{crielaard2022refining}).

In specific scenarios, existing sub-models, or their components, might be sourced from external parties or experts involved in the MMS development process (see, e.g., \citep{cuppen2021participatory}). The MMS and the embedded CLDs within the sub-models provide a mechanism to evaluate the feasibility of repurposing these existing models. They also facilitate discussions around necessary modifications for model integration, including considerations of model ontologies \citep{voinov2013integronsters}). 

Future work needs to focus on the uncertainties introduced through the functional decomposition of a policy problem in interacting sub-models, e.g., related to how uncertainties propagate among sub-models when exchanging information \citep{voinov2013integronsters}. A series of scenarios concerning how information is processed when exchanged among sub-models (e.g., via different mathematical expressions or if-then-else statements) can help identify the domain knowledge required to capture such uncertainty. Scenarios can also include the occurrence of exogenous or endogenous events that trigger the need for particular sub-models (e.g., when a pandemic occurs, an epidemiological model is required) \citep{guivarch2017scenario, yilmaz2007requirements} \citep{guivarch2017scenario}.

Additionally, when the selected paradigm for a sub-model is System Dynamics Modeling (SDM), the portion of the CLD corresponding to the sub-model can be quantified using existing methodologies such as \citep{ford1998expert, mooy2001quantification, crielaard2022refining}. For example, Crielaard et al. \citep{crielaard2022refining} show how a CLD enriched with additional domain knowledge (e.g., functional relationships describing causal links mathematically) can be systematically translated into an SDM. CLDs can also provide useful insights to guide sub-model development when the chosen paradigm is Agent-Based Modeling (ABM) \citep{polhill2021using, abbott2017complex}. This is particularly true when CLDs contain factors, relationships, and feedback loops and include information concerning governance scale intervals acquired through the MMS development process. When this information is available, CLDs elucidate relationships and feedback loops among factors within the same governance scales (termed single-level causation) and those extending across governance scales (termed multi-level causation). This insight proves invaluable for informing the development of ABMs, given that this modeling paradigm is particularly adept at capturing multi-level causation while also accommodating instances of single-level causation \citep{antosz2022sensemaking,gilbert2002varieties, conte2014minding}. However, while methodologies have been proposed to develop ABMs based on other causal mapping techniques such as (fuzzy) cognitive maps \citep{elsawah2015methodology, mehryar2019individual, barbrook2022systems} clearly laid out methodologies are missing for developing ABMs based on CLDs. Consequently, additional research is necessary to adapt and enhance current methodologies for ABM development, exemplified by \citep{bharwani2015, ghorbani_structuring_2015, nespeca_methodology_2023}, to facilitate the construction of ABMs rooted in CLDs. For example, Nespeca et al. \citep{nespeca_methodology_2023} introduce a methodology to develop ABMs that relies on a conceptual framework tailored to a policy problem to guide the data analysis and model development process. Future research should focus on informing the development of such a conceptual framework based on the CLDs corresponding to the sub-models to which an ABM was assigned. Such a development would enable modelers to fully capitalize on the results of an MMS design to inform the sub-models’ development both in the case of SDM and ABM modeling paradigms. 

With time, even for novel policy problems, several models studying policy implications in different scales, perspectives, and possibly relying on different modeling paradigms are likely to be developed in parallel by many actors (as was the case during the COVID-19 pandemic). Ignoring this richness in perspectives misses the opportunity to ‘tap’ into such an evolving landscape of models, their developers, and users (which to some extent also represent the evolving landscape of knowledge regarding the novel policy problem) \citep{bollinger2015,bollinger2018multi}. At this stage, an integrated modeling approach focusing on coupling and fostering the co-evolution of existing models developed by external modeling teams becomes relevant \citep{laniak2013integrated, tucker2022csdms}. 
As an alternative, the integral modeling approach outlined in this study could be adapted to address this challenge by bringing together different teams of modelers and stakeholders to reverse-engineer a conceptual understanding of existing computational models, find relationships among them, study their compatibility, and outline the steps necessary to adjust the existing models to enable their coupling. To this end, further research is required to adapt the procedure proposed in this study to enable modelers and experts to infer conceptual models from their computational counterparts and assess whether different computational models can be coupled and how. It must be noted that such an approach is integral and not integrated. In fact, despite utilizing models initially developed by external sources, the integration process takes place at the conceptual level rather than at the computational level, aligning with the principles of integral modeling.

\section*{Declaration of Competing Interest} 
The authors declare that they do not have any recognized conflicting financial interests or personal connections that might seem to have affected the research presented in this article. 

\section*{Data availability}
Data will be provided upon request.


\appendix

\section{Design choices and requirements for Multi-Modeling Structures} 
\label{SecA1}
\renewcommand{\thefigure}{A.\arabic{figure}}
\renewcommand{\thetable}{A.\arabic{table}}

In this appendix, we discuss literature from the fields of multiscale modeling and multiparadigm (or hybrid) modeling. For each field, we derive a list of key design choices to be made by modelers when developing Multi-Modeling Structures (MMSs). We then infer the domain knowledge from experts required to inform the modeler in making such key design choices (i.e., domain knowledge requirements).

Domain knowledge is defined as the expertise and understanding that particular actors have in a given field such as medicine or psychology which may be necessary to develop computational models that draw from such fields of expertise. Such knowledge is distinguished from computational modeling knowledge which is the methodological expertise required to make key design choices informed by the expert’s domain knowledge to develop computational models that are fit for a specific purpose \citep{edmonds2019different}. 

\subsection{Multi-scale modeling}

Complex multi-domain dynamics can span a wide range of temporal, spatial, or governance scales \citep{van_delden_integration_2007, verburg_multi-scale_2008,  agostinelli2022great}. Individual models trying to capture such a wide range of scales can require considerable computational costs, which can make their use unfeasible \citep{Brandt_multiscale_2001}. When multi-scale modeling is employed, the phenomenon (or complex dynamic) of interest is functionally decomposed into (a) sub-phenomena representing only the essential processes occurring at different sub-scales (e.g. spatial or temporal) and (b) the key interactions among the sub-phenomena \citep{ingram_classification_2004, yang_ontological_2009}. For example, \cite{wang2022multiscale} decompose virological and epidemiological processes at different spatial scales (the host and the population) and temporal scales (fast and slow processes) to provide policy support regarding pharmaceutical interventions. Relying on such functional divisions entails approximations that are likely to reduce accuracy, but at the same time, make the simulation of the considered multi-scale phenomenon feasible \citep{chopard2014framework}.

\subsubsection{Design choices for multiscale modeling}
\begin{figure*} [h!]
    \centering
    \includegraphics[width=0.8\textwidth]{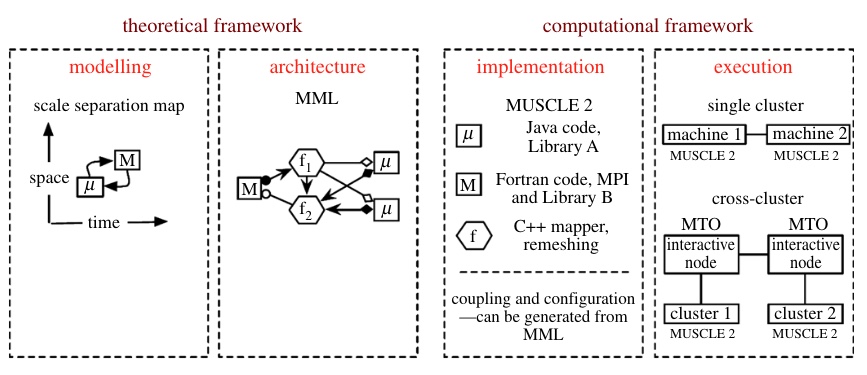} 
    \caption{Overview of the Multiscale Modeling and Simulation Framework (MMSF) \cite[p. 721]{borgdorff2014distributed}. First, develop a Scale Separation Map (SSM) (e.g., with a model $\mu$  at the macro level and model µ at the micro level). Second, translate the scale separation into a modeling architecture (or Multi-Modeling Structure - MMS) by using the Multiscale Modeling Language (MML) that enables to precisely define the way information is exchanged among sub-models e.g. through the mappers $f_1$ and $f_2$. Third, implement the architecture captured through the MML as software through the Multiscale Coupling Library and Environment (MUSCLE). Fourth, execute it on a single machine, a cluster of machines, or multiple clusters through the MUSCLE Transport Overlay (MTO).} 
    \label{fig:mmsf}
\end{figure*}

A few methodological studies propose domain-independent conceptual frameworks for multiscale modeling \citep{ingram_classification_2004, yang_ontological_2009, weinan_heterogeneous_2007, chopard2014framework}. In this study, we refer to one of such frameworks, namely the Multiscale Modeling and Simulation Framework (MMSF) introduced by \citep{borgdorff2013foundations, chopard2014framework, borgdorff2014distributed} and shown in Figure \ref{fig:mmsf}.

\begin{figure*} [h!]
    \centering
    \includegraphics[width=0.7\textwidth]{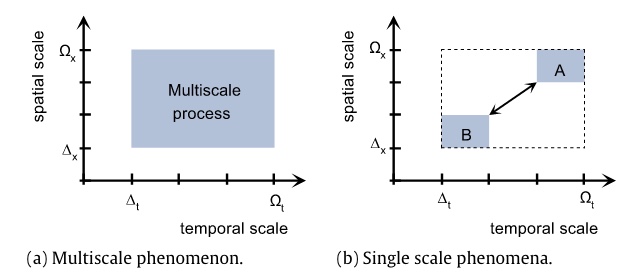} 
    \caption{Scale Separation Map: Functional decomposition of a multiscale phenomenon (or process) in single scale phenomena (represented by the rectangles A and B) and their cross-scale interactions (represented by the bi-directional arrow). From \cite[p. 470]{borgdorff2013foundations}.}
    \label{fig: scale-separation-map}
\end{figure*}

According to the MMSF, there are four key steps in the development of a multiscale model. First, a \textit{scale separation} is carried out and captured through a Scale Separation Map (SSM) (cf. Figure \ref{fig: scale-separation-map}). The SSM illustrates the functional decomposition of a multi-scale phenomenon in sub-phenomena at smaller-scale intervals and their cross-scale interactions. By scale interval, we mean a set of two scales or values representing respectively the smallest and the largest scales that are relevant for the considered phenomenon \citep{borgdorff2013foundations}. These scales are assigned within a specific dimension such as time, space, or governance level \citep{chopard2014framework}. For example, the scale interval of a sub-phenomenon in the temporal dimension is constituted of the time step with which the phenomenon is studied (smallest scale) and of the considered duration of the phenomenon (largest scale). 

Second, the architecture of a multiscale model is designed through the Multiscale Modeling Language (MML) \citep{falcone2010mml}. The architecture includes the sub-models and their interactions or coupling (an architecture corresponds to a Multi-Modeling Structure or MMS as discussed in the introduction). The sub-models correspond to the sub-phenomena obtained through scale separation. The interactions among sub-models specify \textit{when} the sub-models exchange information and \textit{how} such information is exchanged and processed during the exchange among sub-models. With respect to \textit{when} information is exchanged, one sub-model may provide another with the initial conditions required to begin the simulation, or with boundary conditions that the second model requires throughout its simulation. These interactions can be captured through the MML and take the form of different ”coupling templates” depending on when in their execution the models share and receive information \citep{chopard2014framework}. 
Further, the MML enables to capture \textit{how} information is exchanged both in terms of (a) the topology of interactions among the sub-models (capturing information flows among the sub-models) and (b) the scale bridging operations way information is processed when transferred among different models at different scales (here simply referred as operations) \citep{chopard2014framework}. When the information is exchanged among models that have different scales in space or governance levels, information aggregation or dis-aggregation operations are required. 
The topology of interactions among sub-models represents the structure of information flows among sub-models capturing for example whether the information is exchanged in a one-to-one, one-to-many, many-to-one, or many-to-many manner. 

\begin{figure*} [t!]
    \centering
    \includegraphics[width=0.85\textwidth]{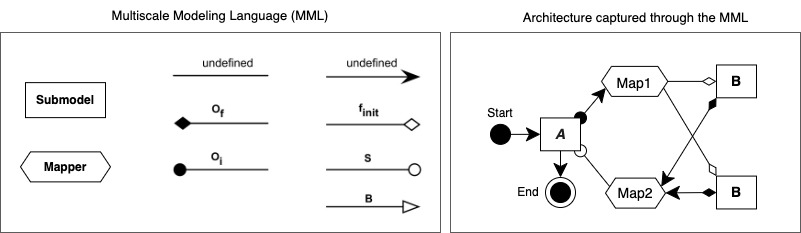} 
    \caption{Multiscale Modelling Language or MML (left) and its application to capture the architecture of a multiscale model based on the scale separation from Figure \ref{fig: scale-separation-map} (right). The MML (left) captures the topology of interactions among sub-models, the coupling templates, and the operations. The topology of interactions representing the information flow among sub-models is captured through the arrows and their orientation. The coupling templates are represented by the symbols adopted for the beginning and tip of the arrows. Specifically, Of stands for final output, the output of the model at the end of its simulation. Oi is an Output exchanged during the simulation at the time step i. Finit shows that the input received by a sub-model defines the initial conditions of the sub-model. S and B represent an input the sub-model can receive during its simulation concerning a Boundary or State condition. The operations are represented by the Mappers. Mappers receive information from sub-model(s), combine and process such information, and transfer the result to other sub-model(s). The example of an architecture (in this article this is termed Multi-Modeling Structure or MMS) captured through the MML (right) shows how the sub-phenomena A and B obtained through scale separation (cf. Figure \ref{fig: scale-separation-map}) represent the sub-models that can be coupled in an architecture. Here, Model A begins its first simulation step, produces the output Oi, and then transfers it to the mapper Map1. Then, Map1 carries out a scale bridging operation on Oi and transfers the resulting information to two instances of Model B. This incoming information is used by the two instances of Model B to set their initial conditions (Finit). These instances then run until their simulation is completed and transfer their final output Of to the mapper Map2. Map2 combines and processes the information received through another scale bridging operation and transfers the results to Model A. Model A uses the information received to set its state condition S before proceeding with another simulation step. This cyclic simulation repeats until the simulation of model A is completed. Based on \citep{falcone2010mml, borgdorff2013foundations}.}
    \label{fig:mml}
\end{figure*}

Third, the architecture captured through the MML can be directly implemented in the Multiscale Coupling Library and Environment (MUSCLE). MUSCLE enables combining different sub-models (possibly written with different programming languages) and running simulations with them as if they were one single model. Fourth, once the architecture is implemented in MUSCLE, simulations can be executed both on a single machine and on clusters or supercomputers \citep{chopard2014framework}. 

\subsubsection{Domain knowledge requirements for multiscale modeling}
In the following, we discuss the domain knowledge required to inform the key design choices introduced in the previous section.

First, to design the SSM it is necessary to (a) functionally decompose the multiscale phenomenon in sub-phenomena each to be captured by a sub-model and (b) identify the cross-scale interactions among such sub-models. In terms of domain knowledge, this requires a clear definition of the multiscale phenomenon to be modeled. This topic can be sketched by researchers and modelers, but it is ultimately those who have access to domain knowledge (i.e., the experts) that should be involved in clearly defining the system and phenomenon of interest. Next, it is necessary to identify the multimodel space, i.e., the key dimensions such as time, space, and governance levels within which the scale intervals of the phenomenon and sub-phenomena are defined \citep{chopard2014framework}. In order to carry out scale separation, the considered phenomenon must be functionally decomposed in the most crucial sub-phenomena that are sufficiently independent from one another so that scale separation can be carried out. This separation can be obtained based on domain knowledge about the key phenomena and their associated scale. Finally, the \textit{scale intervals} for each of the sub-phenomena must be defined in the multi-model space. In sum, the \textit{multiscale phenomenon}, \textit{multi-model space}, \textit{functional decomposition in sub-phenomena}, and \textit{scale intervals} are domain knowledge requirements to carry out the SSM.

Second, it is key to design the architecture of the multiscale model, which entails characterizing when and how information is exchanged among sub-models given their interactions. 
The \textit{temporal scales intervals} enable to fully capture when sub-models exchange information \citep{chopard2014framework}. Given the temporal scales were already captured in the scale separation map, no additional domain knowledge is required to inform this design choice. Next, the \textit{spatial or governance scale intervals} of the sub-models provide the means to establish whether the information exchanged among the sub-models needs to be aggregated or disaggregated through scale bridging operations and techniques. While the difference in scales among interacting sub-models indicates the need for scale bridging operations, such operations need to be discussed ad hoc and require domain knowledge. As such, scale bridging operations constitute a domain knowledge requirement to be discussed with experts. Next, (a) the \textit{topology of the interactions} among sub-models and (b) the \textit{operations} that may be required to combine information from different models before re-directing it to other models, constitute two additional domain knowledge requirements that cannot be addressed simply based on scale.

The last two steps, namely, implementation and execution rely on the competence and ability of modelers to implement the multiscale modeling architecture as software (e.g., in MUSCLE) and to execute simulations on a single or on cluster(s) of machines. However, no additional domain knowledge is required from experts at this step.

Table \ref{tab:req-multi-scale} summarizes the domain knowledge requirements found for each of the four steps of the MMSF framework.

\begin{table*}[!h]
\centering
\caption{Key design choices in multiscale modeling, and the domain knowledge requirements for each of the choices.}\label{tab:req-multi-scale}
\begin{tabular*}{\textwidth}{lp{0.64\textwidth}}
\toprule%
\textbf{Key design choices} & \textbf{Domain knowledge requirements} \\
\midrule
Scale Separation & Multiscale phenomenon of interest, multi-model space (i.e. the relevant dimensions considered for scale separation such as time, space, and governance level), functional decomposition in sub-phenomena, scale intervals of the sub-phenomena in the multi-model space. \\
\midrule

Architecture & Topology of the interactions or information exchange among sub-models (1:1, 1:M, N:M), Scale bridging operations required to manage and combine information when exchanged among sub-models at different scales.\\
\midrule

Implementation & N.A. \\ \midrule

Execution & N.A. \\ 

\bottomrule
\end{tabular*}
\end{table*}

\subsection{Multi-paradigm modeling}

Similarly to multiscale modeling, multi-paradigm modeling entails the combined use of multiple models to capture a particular phenomenon or achieve a given objective of interest. While there is no consensus on the definition of Multi-Paradigm Modeling (MPM) \citep{mustafee2017purpose}, a common denominator for this research field is the integrated use of multiple modeling formalisms or paradigms \citep{zeigler1986multifaceted, zeigler2000theory, vangheluwe2002introduction, borshchev2004system, mykoniatis2020modeling, barivsic2022multi, kwon2024hybrid}. A modeling formalism is an approach that modelers can use to build models of a system \citep{zeigler1986multifaceted, zeigler2000theory}. Examples of modeling formalisms include Discrete Event Simulation (DES), SDM, and ABM. Modeling formalisms typically present different sets of hypotheses concerning the ”underlying sources of dynamics in a system” \citep{lorenz2006towards}. For example, in the case of SDMs, positive and negative feedback loops among aggregate variables and accumulations of flows are considered as the sources of dynamics in a system \citep{sterman2000business, scholl2001agent}. Conversely, in ABMs the interactions among individual agents and their environment (micro-macro-micro) are considered as the sources of dynamics \citep{epstein1996growing, Bonabeau2002}.

\subsubsection{Design choices for multi-paradigm modeling}

Many frameworks have been introduced that propose approaches for the development of multi-paradigm and hybrid models including \citep{atom3,  chahal2013conceptual, borshchev2014multi, mykoniatis2020modeling, jones2021aiding}. Some of these frameworks such as \cite{chahal2013conceptual} and \cite{mykoniatis2020modeling} introduce and characterize key design choices in multiparadigm modeling: problem identification and decomposition, choice of modeling paradigms, and identification of the interactions among models.

First, problem identification and decomposition entail the definition of the policy problem considered, the associated modeling objective, and the decomposition of the modeling objective in sub-objectives each to be captured by a single model. At this stage, also project constraints are discussed such as time, budget, and data availability \citep{mykoniatis2020modeling}.

Second, the choice of a modeling formalism is carried out for each of the submodels. Several MPM methodological studies provide criteria for the choice of an adequate modeling formalism. \cite{chahal2013conceptual} and \cite{lorenz2006towards} stress that the development of valuable MPMs depends on the alignment of the modeling formalism with the modeling subobjective (or problem) and the context (or system) considered. The properties of a sub-objective that are relevant for the choice of a modeling formalism can be e.g., whether the focus is placed on tracking individual or aggregate behavior \citep{sterman2000business, kelly2013selecting}, the strategic or operational nature of the considered problem, and the modeling purpose such as predicting or explaining a particular system’s behavior of interest \citep{lorenz2006towards, edmonds2019different}. Another criterion related to the problem or modeling objective is feasibility given the available time. ABMs tend to entail much higher computational costs and require longer development times given their ability to capture more details compared with e.g., SDM \citep{rahmandad2008heterogeneity}. As such, in the case of urgent situations such as crises, developing ABMs could be less feasible. Another factor is the availability of already existing ABMs (e.g., through online libraries) that were previously verified and validated by the scientific community and that can be adapted and combined in a modular manner. Such availability can play a key role in making the use of ABMs more feasible within shorter time frames \citep{squazzoni2020computational}.

Next, properties of the system that are relevant for the choice of a modeling formalism include the heterogeneity in attributes of individual actors/agents \citep{dyke1998agent, Bonabeau2002, rahmandad2008heterogeneity}, the complexity of their individual behavior (e.g. learning and adaptation) \citep{Bonabeau2002}, their patterns or topology of interactions (e.g. non-perfect mixing or clustering) \citep{Bonabeau2002, rahmandad2008heterogeneity}, the continuous or discrete behavior of the system \citep{wilson1998resolving, zeigler2000theory, shnerb2000importance, cellier2006continuous}, and the level of detail and uncertainty at which data and knowledge are available [18, 52]. Table \ref{tab:req-multi-par} lists some of the criteria that are relevant for the choice of a modeling paradigm or formalism and examples in which each criterion is relevant in isolation or in combination with the other criteria. This table is not meant to be a complete overview of all criteria required to inform the choice of a modeling paradigm. Rather, we aim to show some examples of criteria related to the problem and system of interest that are relevant when choosing a modeling paradigm.

\begin{table*}[!h]
\centering
\caption{Summary of the criteria for the choice of a modeling paradigm, examples in which such criteria were relevant, and suggestions on how each of the criteria can inform the choice of an ABM or SDM modeling paradigm.}\label{tab:req-multi-par}
\begin{tabular*}{\linewidth}{p{0.2\linewidth} p{0.2\linewidth} p{0.17\linewidth} p{0.17\linewidth} p{0.1\linewidth}}

\toprule%

\textbf{Criteria for formalism selection} & \textbf{Examples} & \textbf{Formalism: SDM} & \textbf{Formalism: ABM} &  \textbf{Source} \\ \midrule 

Heterogeneity in attributes & Superspreading events \citep{lloyd2005superspreading} & Yes, but limited to a few Heterogeneous sub-groups & Yes & \citep{dyke1998agent, Bonabeau2002, rahmandad2008heterogeneity} \\ \midrule 

Mixing (patterns in space or topology of interactions) & 
Contact networks in epidemiology \citep{bansal2007individual} & Yes, if clusters of interactions can be subdivided into groups that are stable over time & Yes & \citep{Bonabeau2002, rahmandad2008heterogeneity} \\ \midrule 

Complexity of behavior (Adaptation and Learning) & Fear acquisition and extinction in combined virus and fear dynamics \citep{epstein2014agent_zero} & Yes, if heterogeneity and mixing do not play a major role & Yes & \cite{Bonabeau2002, fenichel2011adaptive} \\ \midrule 

Discrete system's behavior & Extinction of biological species  \citep{wilson1998resolving} & Yes, if knowledge on when to "impose" it & Yes &  \citep{shnerb2000importance,cellier2006continuous,zeigler2000theory}\\ \midrule 

Aggregation level, availability, and uncertainty of Knowledge and Data & Knowledge is available at individual or aggregate level & Yes, if available at macro level & Yes, if available at micro (and possibly macro) level & \citep{lorenz2006towards, Bonabeau2002, rahmandad2008heterogeneity} \\ 

\bottomrule
\end{tabular*}
\end{table*}

Third, the identification of the interactions among the sub-objectives (and the respective sub-models) is carried out. To map and characterize such interactions, \citep{chahal2013conceptual} suggest (a) identifying the interaction points (ie. possible points for information exchange among the sub-models), (b) finding relationships among the interaction points (i.e. information flows among the sub-models), and (c) identifying the mode of interactions (i.e. the way information is processed when the sub-models change information). The interaction points are identified by considering variables that are placed in two different sub-models but present the same meaning (though may have different levels of aggregation) and the models’ input and outputs \citep{chahal2013conceptual}. \cite{mykoniatis2020modeling} adds that inputs and outputs of the model can be further classified depending on whether the models exchange information before, during, or after a model is run. This information is the same as that captured by the coupling templates in the MML introduced by \citep{falcone2010mml, chopard2014framework} and presented in the section ”Design choices for multiscale modeling”.

Next, the relationships among interaction points are studied. Such relationships can be of the type ”value assignment” or ”impact statement” \citep{mykoniatis2020modeling}. Value assignment relationships can be the direct replacement of values of variables that are present in both of the interacting models when the variables have the same meaning and the same level of aggregation. In other cases, value replacement requires the dis-aggregation or aggregation of variables when the variables have the same meaning but different levels of aggregation (this can be addressed through scale bridging techniques from multiscale modeling). Another form of value replacement relationship is that of \textit{causal relationships} expressed through predefined mathematical expressions linking one variable to the other. Impact statements are related to relationships describing the influence that a model has on another model which cannot be expressed as a value assignment. For example, when a particular threshold has been reached say in an SDM model, then new agents need to be introduced in an ABM model. These statements also depend on the modeling paradigms adopted for each of the sub-models that present an interaction \citep{mykoniatis2020modeling}.

Finally, the modes of interactions capturing the computational nature of the interactions among sub-models are identified. The modes of interactions can be cyclic (the models are run one after the other) or in parallel (the models exchange information during run time) \citep{chahal2013conceptual}. Designing the modes of interactions requires information regarding the level of aggregation or detail required for the sub-models \citep{lorenz2006towards,kelly2013selecting}.

\subsubsection{Domain knowledge requirements for multi-paradigm modeling}
The domain knowledge required to inform each of the design choices for MPM is discussed in the following.

First, with respect to problem identification and decomposition, it is key to define an objective for the modeling exercise. Domain knowledge is required to identify the policy problem to be addressed including the considered case and the objective of the model e.g. to support policy formulation or evaluation \citep{gilbert2018computational}. Additionally, functionally decomposing the modeling objective relies on domain knowledge to identify what are the key and potentially interrelated sub-objectives to be considered.

Second, choosing a modeling paradigm also requires domain knowledge. First, information regarding the type of problem to be addressed, and project constraints (e.g. time availability) are key in informing the choice of a modeling paradigm. Second, properties of the considered system in which the problem needs to be addressed also play a key role in the choice of a modeling paradigm. Such properties include heterogeneity in the factors to be considered (in attributes among the different actors considered in the model), mixing as in the patterns of interactions among actors, and with their environment, the complexity of behavior considered (e.g., adaptation based on context), discreteness in system’s state and behavior, availability of and uncertainty in data and knowledge at different levels of abstraction (including the availability of previously developed and validated models).

Third, to map the interactions among the models, the interaction points, relationships among such interaction points, and the levels of aggregation required for the sub-models need to be identified based on domain knowledge.

Table \ref{tab:req-multi-paradigm} summarizes the design choices related to MPM, and the associated domain knowledge required to inform such choices.

\begin{table*}[h]
\centering
\caption{Key choices in multi-paradigm modeling, and the domain knowledge requirements for each of the choices.}\label{tab:req-multi-paradigm}
\begin{tabular*}{\linewidth}{p{0.3\textwidth} p{0.65\textwidth}}

\toprule%
\textbf{Key design choices} & \textbf{Domain knowledge requirements} \\
\midrule
Problem Identification and Decomposition & Policy problem, modeling objective, functional decomposition in key and potentially interacting sub-objectives. \\ \midrule

Modeling paradigm selection & Properties of the problem considered: nature of the problem, project constraints, modelling objective. Properties of the system: Heterogeneity, mixing, discreteness, complexity of behavior, availability of data and knowledge. \\ \midrule

Identification of interactions
& interaction points, relationships among interaction points (value assignment and impact statements), mode of interaction among the sub-models (cyclic or in parallel)\\ 

\bottomrule
\end{tabular*}
\end{table*}

\subsection{Multi-modeling: Multiscale and Multi-paradigm modeling}
Both MSM and MPM belong to the wider field of multi-modeling, which entails the combined use of different models to capture a phenomenon of interest. While MSM focuses on approximating a phenomenon of interest through the use of multiple submodels at different scales \citep{chopard2014framework}, MPM is centered on the adoption of multiple modeling formalisms that are adequate for different sub-components of the problem and system considered \citep{chahal2013conceptual}. The two approaches are not mutually exclusive as a multi-model can include sub-models at different scales, captured through multiple modeling paradigms. In the following, the design choices and the associated domain knowledge requirements for multi-modeling are discussed based on the design choices and domain knowledge requirements identified respectively for MPM and MSM (cf. Tables \ref{tab:req-multi-scale} and \ref{tab:req-multi-paradigm}).

\subsubsection{Design choices for Multi-modeling}

First, it is key to establish the purpose of the modeling endeavor. In the case of MSM, the focus is placed on capturing and understanding the phenomenon to be studied \citep{chopard2014framework}. Conversely, MPM stresses more generally the need to identify the problem and the associated objective of the modeling exercise \citep{chahal2013conceptual, mykoniatis2020modeling}. Given that this article aims at developing multi-models for policy support, the focus is here placed on identifying the policy problem to be captured and the associated modeling objective e.g., to explore future policy implications during policy formulation, or to explain the impact of policies in the past to inform policy evaluation \citep{gilbert2018computational}.

Second, both in MSM and MPM it is essential to study if and how to functionally decompose the objective or phenomenon of interest in sub-models. The decomposition involves both the division in models at different scales and/or relying on different modeling paradigms. The two choices are not independent as the selection of a modeling formalism is often influenced by the required level of resolution or abstraction \citep{zeigler1986multifaceted, vangheluwe2002introduction, borshchev2004system, mykoniatis2020modeling}, which is associated with the considered scale \citep{borgdorff2013foundations, chopard2014framework}. As such, at this stage, joint considerations about the functional decomposition in sub-models, the assigned modeling paradigm, and scale separation are required.

Third, to complete the design of the architecture of a multi-model or multimodeling structure (MMS) it is necessary to identify the interactions among the sub-models. In the following, we refer to the concept of interactions as described in MSM literature but also draw a comparison with MPM (cf. Tables \ref{tab:req-multi-scale} and \ref{tab:req-multi-paradigm}). The interactions among sub-models include (a) the coupling templates describing when the models interact, (b) the topology o of interactions describing information flows among the sub-models (defined as relationships in MPM), and (c) the operations carried out to process information when it is transferred among different models (defined as relationships types in MPM). From what we learned from both MSM and MPM literature, the different types of operations include value transfer, scale bridging operations (including aggregation/disaggregation), mathematical expressions representing causal relationships, impact statements (e.g. if, then, else statements), and combinations of such different types of operations. 
The MML enables modelers to capture an MMS including the sub-models and their interactions (cf. Figure A.3). When used also for MPM, the MML should include extra annotations to show the modeling paradigm chosen for the sub-models.

\subsubsection{Domain knowledge requirements}

First, the identification of the policy and modeling objective requires domain knowledge from experts about the policy problem to be addressed. Next, the modeling objective can be discussed between modelers and experts also considering feasibility given time, budget, and data availability constraints (Cf. Table \ref{tab:req-multi-paradigm}). As such, domain knowledge concerning project constraints and the modeling objective is also required at this stage.

Second, the scale separation requires the definition of a multi-model space, the decomposition in sub-models, and the identification of the scale intervals corresponding to each sub-model defined within the multi-model space (Cf. Table \ref{tab:req-multi-scale}). The selection of and further decomposition in different modeling paradigms requires the characteristics of the considered policy problem (already available from the previous step), and the level for aggregation or detail required corresponding to the scale intervals defined in the functional decomposition in scales (Cf. Table \ref{tab:req-multi-paradigm} and \ref{tab:req-multi-scale}). Additionally, the functional decomposition in sub-models also requires domain knowledge associated with the particular system considered including its level of heterogeneity, mixing, discreteness, complexity of behavior, and the availability of data and knowledge (Cf. Table \ref{tab:req-multi-paradigm}).

Third, the \textit{definition of the sub-models interactions} requires their scale intervals (in space or governance level and time). This knowledge is already available from the scale decomposition. Next, the topology of interactions among sub-models constitutes two additional domain knowledge requirements. The operations needed to process and combine information when it is transferred among sub-models are also a requirement. Such operations can be of different types with different requirements. In the case of value transfer, the same exact value is exchanged from one model to the other. These types of operations do not require additional domain knowledge. However, other operations such as scale bridging, causal relationships (computed through mathematical expressions), or impact statements (e.g., if, then, else operations) require further domain knowledge from experts and, as such, constitute an additional requirement (Cf. Table \ref{tab:req-multi-paradigm} and \ref{tab:req-multi-scale}).

Table \ref{tab:req-multi-modeling} in the section ”Results” summarizes the design choices and the associated domain knowledge requirements for multi-modeling found in the literature as discussed in this appendix.

 \bibliographystyle{elsarticle-num} 
 \bibliography{references.bib}





\end{document}